\documentstyle[aps,epsf,eqsecnum,twocolumn]{revtex}
\input epsf.sty
\begin{document}
\draft
\flushbottom
\twocolumn[
\hsize\textwidth\columnwidth\hsize\csname @twocolumnfalse\endcsname

\title{Luttinger stripes in antiferromagnets}
\author{A. H. Castro Neto}

\address{Department of Physics,
University of California,
Riverside, CA, 92521}
\date{\today}
\maketitle
\tightenlines
\widetext
\advance\leftskip by 57pt
\advance\rightskip by 57pt

\begin{abstract}
We propose a model for the physics of stripes in antiferromagnets
in which the stripes are described by Luttinger liquids hybridized
with antiferromagnetic domains. Using bosonization techniques
we study the model in the limit where the magnetic
correlation length is larger than the inter-stripe distance and
propose an explanation for the commensurate-incommensurate
phase transition seen in neutron scattering in the underdoped
regime of La$_{2-x}$ Sr$_x$ Cu O$_4$.
The explanation is based on a phase to
anti-phase domain transition in the spin configuration which
is associated with the transverse motion of the stripes.
Using a non-linear $\sigma$ model to describe
the antiferromagnetic regions we conjecture the crystalization of
the stripes in the magnetically ordered phase.
\end{abstract}
\pacs{PACS numbers: 75.25.+z, 74.72.-h, 74.72.Dn, 73.20.Dx, 71.27.+a}

]
\narrowtext
\tightenlines

\section{Introduction}

In recent years we have seen  tremendous development in
the production of new materials which have a phase diagram
where an insulating magnetic state is very close to a metallic
or superconducting state. The most famous examples are
the high temperature superconductors such as La$_{2-x}$ Sr$_x$
Cu O$_4$, Bi$_2$ Sr$_2$ Ca Cu$_2$ O$_{8+\delta}$ and Y Ba$_2$ Cu O$_{6+x}$
which are formed by layers of Cu O$_2$. Their phase diagram is well-known.
At zero doping the
ground state is insulating and
antiferromagnetic. Antiferromagnetism is suppressed very rapidly with doping
and it is followed by
a spin-glass phase and a superconducting
state with high critical temperature. There are many anomalies in these
systems when compared to a normal Fermi liquid, which have been a source
of a lot of controversy in the last few years.

\begin{figure}
\epsfysize5cm
\hspace{1cm}
\epsfbox{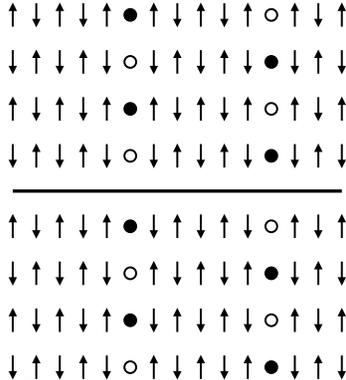}
\caption{Geometry of the problem represented by vertical
stripes (shown with black and white circles)
inside of an antiferromagnet (which is represented by
$\uparrow$ and $\downarrow$). In the figure on the
top we show a {\it phase domain} and on the bottom we show
an {\it anti-phase domain}. Observe the orientation of the
spins.}
\label{stripe}
\end{figure}

In this paper we will introduce a model for the discussion of
these systems in the presence of
segregation of charge into one-dimensional
domain walls or stripes.
The geometry of the
problem is shown in Fig.(\ref{stripe}). The problem of formation
of domain walls and stripes has been discussed for quite some time
in terms of the proximity of these materials to phase separation \cite{ps}.
Macroscopic phase separation has been observed in
La$_{2}$ Cu O$_{4+\delta}$ \cite{hammel}, and stripes have been seen
experimentally in La$_{2-x}$ Sr$_x$ Ni O$_{4+y}$ \cite{nickel} in
many different experiments including electron diffraction. Magnetic
susceptibility measurements \cite{suscep}, nuclear quadrupole resonance
\cite{nqr} and muon spin resonance \cite{msr}
seem to indicate formation of domains in La$_{2-x}$ Sr$_x$ Cu O$_{4}$
in certain ranges of doping and temperature.
More recently a direct evidence for stripe formation was given
in neutron scattering by Tranquada {\it et al.} in
La$_{1.6-x}$ Nd$_{0.4}$ Sr$_x$ Cu O$_{4}$ \cite{tranquada}.
In these experiments Nd was used to stabilize the low temperature
tetragonal phase (LTT) which has a tilt of the oxygen octahedra which is
favorable for stripe pinning. These experiments seem to be
indicative of an inhomogeneous ground state.
Moreover, the existence of stripes in cuprates and ladders
have been confirmed in many different computer simulations
of $t-J$ and Hubbard models \cite{simul}. In these
models the formation of domains is associated with the
formation of pairs of holes and segregation from the magnetic
regions. In the simulations only short range interactions are
included and it is not clear whether long range forces are
important in order to stabilize the stripe structures, especially
because the materials under consideration are poor metals and
screening is not guaranteed. Actually
the idea of a frustrated phase separation due to long range forces
as discussed by Emery and Kivelson \cite{ek} has been very
successful in explaining many of the important features of
high temperature cuprates. Their theory should be contrasted with
theories where it is Fermi surface effects which drive 
the formation of inhomogeneous states \cite{cdw}.

Here we look at this problem from a
more phenomenologic point of view and do not ask for the origin
of the instability towards stripe formation but rather try to
develop the consequences which come from it.
One of the main features is that the charge of the dopants
goes to specific quasi-one dimensional regions of the system which
are surrounded by {\it undoped} quasi-two dimensional regions.
In this way the problem splits
into two independent parts which hybridize with each other: a
quasi-one-dimensional interacting electron gas and a quasi-two dimensional
magnetic electron system.
A lot is known about these
two systems separately. The one-dimensional system has been studied
via bosonization and renormalization group techniques, and it is a
very mature field \cite{review}.
The field of itinerant magnetism has seen a lot of development especially
from the point of view of renormalization group \cite{hertz}
and effective spin models \cite{chn}.
The objective of this work is to try to link these two approaches in
order to gain insight into the physics of these materials.

In section II we propose a model with short range forces for
the physics of the stripes. In section III we discuss the
antiferromagnetically ordered region and focus on the effect
of the stripes on the antiferromagnet and vice-versa. We also
compare our results with recent neutron scattering data.
In section IV we study the magnetically disordered phase and
suggest a possible phase diagram for the stripes. Section V
contains our conclusions.

\section{The Model}

\begin{figure}
\epsfysize4cm
\hspace{1cm}
\epsfbox{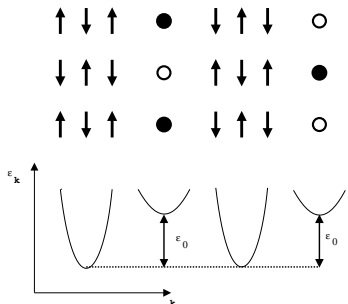}
\caption{Schematic representation of the problem in
terms of coupled chains.}
\label{model}
\end{figure}

Although the basic model for the problem of itinerant holes
in magnets can be quite complicated we are going to simplify
it by making assumptions about the geometric structure of the
charge distribution in the planes of Cu O$_2$.
Instead of starting from a microscopic model
such as a $t-J$ or Hubbard model with possible long-range
interactions we choose to write an effective model which
has already incorporated into it the minimal physical requirements
for the presence of stripes. In this model one assumes that
the regions rich in carriers (which we call stripes)
are separated from the regions
with no carriers (which we call magnetic regions).
This segregation can be introduced if we assume that the energy
of the electrons is locally raised relative to the magnetic regions.
In this picture we look at
the problem as a set of coupled chains where the chains that
``contain" the stripes
have the bottom of the band raised in energy relative to the
magnetic regions (see Fig.(\ref{model})).
In this way the holes tend to segregate into these
chains --- that is, form stripes. The electrons in the stripes 
can hybridize
with the magnetic regions, and this hybridization
gives rise to charge fluctuations. The amount of fluctuations is
controlled by the shift in energy and the hybridization.

For the stripes and the magnetic regions we use a simple Hubbard model.
Our complete model reads,
\begin{eqnarray}
{\cal H} = {\cal H}_H + {\cal H}_S + {\cal H}_I
\label{mod}
\end{eqnarray}
where the magnetic region is described by the Hubbard model
\begin{eqnarray}
{\cal H}_H &=& -t 
\sum_{\langle \vec{r},\vec{r'}\rangle ,\alpha=\uparrow,\downarrow}'
c_{\alpha}^{\dag}(\vec{r}) c_{\alpha}(\vec{r'})
+ c.c.
\nonumber
\\
&+& U \sum'_{\vec{r}} n_{c,\uparrow}(\vec{r}) n_{c,\downarrow}(\vec{r})
\label{init}
\end{eqnarray}
where $\vec{r} = a (n,m)$ labels the sites in
a two-dimensional square lattice ($a$ is the lattice spacing), 
$n$ labels the $x$ direction (perpendicular to the stripe),
$m$ labels the $y$ direction (parallel to the stripe) and the
brackets $\langle,\rangle$ imply sum over nearest neighbor sites.
$c_{\alpha}^{\dag}(\vec{r})$ creates a fermion with spin projection
$\alpha$ at the site $\vec{r}$ in the magnetic region.
$n_{c,\alpha} = c_{\alpha}^{\dag}(\vec{r}) c_{\alpha}(\vec{r})$
is the occupation number at site $\vec{r}$.
The stripes are assumed to be periodically
located at $x= n \ell$
where $n$ is an integer and $\ell=N a$ is the interstripe distance
(the superscript in (\ref{init}) means that the chains at
$n=0,\pm N, \pm 2 N,...$ are
excluded from the sums). The stripe Hamiltonian is also a Hubbard
model that reads,
\begin{eqnarray}
{\cal H}_S &=& -{\tilde t} \sum_{m,n,\alpha = \uparrow,\downarrow}
a^{\dag}_{\alpha}(n N,m) a_{\alpha}(n N,m+1) + c.c.
\nonumber
\\
&+& \epsilon_0
\sum_{m,n,\alpha = \uparrow,\downarrow}  n_{a,\alpha}(n N,m)
\nonumber
\\
&+& \tilde{U} \sum_{m,n} n_{a,\uparrow}(n N,m) n_{a,\downarrow}(n N,m)
\label{hubbard}
\end{eqnarray}
where we are assuming the possibility of mass and vertex
renormalization (${\tilde t} \neq t$, $\tilde{U} \neq U$)
and $\epsilon_0$ is the shift in
energy as it is shown in Fig.(\ref{model}).
$a^{\dag}_{\alpha}(n N,m)$ creates a fermion with spin projection
$\alpha$ on the site $m$ along the stripe located at $n N$.
We assume that the operators $c_{\alpha}(n',m')$ and
$a_{\alpha}(n N,m)$ {\it anti-commute} with each other.
Moreover we assume that after
the stripes are formed the long range Coulomb interaction
within the stripes is screened by the other stripes in the material.
The interaction between the stripes and the magnetic region
is given by the hybridization term,
\begin{eqnarray}
{\cal H}_I &=& \sum_{m,n,\alpha= \uparrow,\downarrow}
V_{m,n} a^{\dag}_{\alpha}(n N,m) \left( c_{\alpha}(n N-1,m) \right.
\nonumber
\\
&+& \left. c_{\alpha}(n N+1,m) \right) + c.c.
\nonumber
\end{eqnarray}
where $V_{n,m}$ is the local hybridization of the stripe with
the magnet region. Observe that when $\tilde{U} = t = 0$ the
problem is equivalent to the one in the Anderson model.
The antiferromagnetic regions behave like $f$ electrons while
the stripes behave like a conduction band. In this sense the
problem becomes very similar to the problem of heavy fermions
in rare earth alloys.
Moreover, observe that if $\epsilon_0 > E_F$
where $E_F$ is the Fermi energy of the magnetic region then the
stripes are going to be completely empty, that is, full of holes
(see Fig.(\ref{model})). This seems to be the case in
La$_{2-x}$ Sr$_x$ Ni O$_{4+y}$ \cite{nickel}.

The main advantage of this Hamiltonian is that it has only short range
interactions and one assumes that all the strong interaction
effects are already renormalized. That is, it is an effective
Hamiltonian for quasi-electrons in the Cu O$_2$ planes. Notice that
the stripes are present even for small values
of the effective parameters which also makes the problem
easier to treat. Moreover, the phenomenologic
parameters $t,U,{\tilde t},\tilde{U},V,\epsilon_0,\ell$ are functions
of doping (and possibly temperature) and they depend intrinsically
on the microscopic details, which are not available. In particular
in the absence of doping (half-filling) one has
$\epsilon_0=0$, $t={\tilde t}=V$ and $U=\tilde{U}$ which leads
to a two-dimensional half-filled Hubbard model.
Naturally the full solution of this problem is of major
complexity. It turns out, however, that this model simplifies
considerably if one considers the physics close to the antiferromagnetic
ordered region.

The closeness to the magnetically ordered region
is measured by the magnetic correlation length, $\xi$.
$\xi$ must be compared with the other characteristic scales in the
problem, which are Fermi wavelength which is proportional to the
lattice spacing, $a$, and the distance between the stripes, $\ell$.
We will assume that the Fermi wavelength of the quasi-electrons
is the smallest scale in the problem
so that we are dealing with a large bandwidth.
The change in the physics of the problem,
from our point of view, comes from the comparison of $\xi$ with
$\ell$. Close to the antiferromagnetic region one expects $\xi > \ell$
and, in particular, in the magnetically ordered phase $\xi$
diverges. In this case large regions of the antiferromagnet are
locked together in a Neel order. In particular, on the scale of
$\xi$ many stripes are found, and it costs a large amount
of energy for the holes to move around. Part of this energy is
compensated by the shift of the bottom of the bands, that is,
$\epsilon_0$ (this is the equivalent of the gain of  Coulomb
energy with condensation into stripes). Therefore, in this
region, the holes are going to be confined to the stripes.
When $\xi$ becomes comparable to or smaller than $\ell$ one expects
the holes to make excursions into the antiferromagnetic regions,
because the order is not robust enough to segregate the holes into
the stripes.
In the next section we discuss the physics of the problem
close to the magnetically ordered state.

\section{$\xi > \ell$, magnetically ordered phase.}

In this section we assume that the
charge fluctuations into the magnetic regions are very rare
because the system is close to magnetic order and large regions
of the magnet are locked into an almost ordered state. As is well
known, the holes
disrupt the order by overturning spins and therefore are expelled
from the magnetic regions into the stripes. Thus, charge fluctuations
are only allowed within the stripe. In terms of our model this implies
that $U>>t$ and $V<<\epsilon_0$. This allows us to do perturbation
theory in $t/U$ and $V/\epsilon_0$.

When $U>>t$ the magnetic regions (which are half-filled)
can be mapped into the Heisenberg model using degenerate
perturbation theory,
\begin{eqnarray}
{\cal H}_H = J \sum'_{\langle \vec{r},\vec{r'} \rangle} 
\vec{S}(\vec{r}) \cdot \vec{S}(\vec{r'})
\label{Heisenberg}
\end{eqnarray}
where $J = 2 t^2/U$ is the exchange constant. In this limit
the electrons in the stripes can only hop virtually into the
magnetic regions. It is easy to show that in this case the
allowed interaction at low energies is an exchange coupling
which can be derived via a Schrieffer-Wolf transformation
for small $V$ \cite{sw}. The hybridization term has the
form of a Kondo coupling,
\begin{eqnarray}
{\cal H}_I &=& J' \sum_{m,n} \vec{S}_e(n N,m) \cdot
\left(\vec{S}(n N+1,m) \right.
\nonumber
\\
&+& \left. \vec{S}(n N-1,m)\right)
\label{coupling}
\end{eqnarray}
where  $\vec{S}_e(n N,m) = \sum_{\alpha,\alpha'} a^{\dag}_{\alpha}(n N,m)
\vec{\sigma}_{\alpha,\alpha'} a_{\alpha'}(n N,m)$ is the electronic
spin in the stripe and $J' \propto \frac{|V|^2}{\epsilon_0}$
is the stripe exchange ($\vec{\sigma}_{\alpha,\alpha'}$ represents
the Pauli matrices). We have obtained a major simplification
now because the charge fluctuations in the magnetic regions are
frozen; only magnetic fluctuations are allowed. Thus, we can focus
only on the spin dynamics.

Since we are assuming that the magnetic correlation length is
larger than the interstripe distance, we can work with just
one stripe at a time, because each stripe sees an effective
magnetic environment. Moreover, in this region, because the
levels of doping are small, the inter-stripe distance is
much larger than the lattice spacing and the connection
between the stripes is weak.
Therefore in what follows we look at the
dynamics of one isolated stripe and assume that all the other
stripes essentially undergo the same dynamics. This also leads to
further simplification in the calculations. In the magnetic
region the stripes do not fluctuate transversally. Therefore
their physics is going to be described in terms of a {\it Luttinger
liquid}, that is, an interacting one-dimensional Fermi gas.
We can therefore use the powerful machinery of bosonization in
order to study their physics \cite{review}.
The fermion operator is written in terms
of right, $R$, and left, $L$, moving electrons as,
\begin{eqnarray}
a_{\alpha}(y) = \psi_{R,\alpha}(y) e^{i k_F y}
+ \psi_{L,\alpha}(y) e^{-i k_F y}
\nonumber
\end{eqnarray}
and these can be bosonized via the transformation
\begin{eqnarray}
\psi_{R,L,\alpha}(y) = \frac{1}{\sqrt{2 \pi \eta}} e^{\pm i \sqrt{\pi}
\phi_{R,L,\alpha}(y)}
\nonumber
\end{eqnarray}
where $\eta$ is a lattice cut-off for the theory in the continuum. 
The bosonic modes
$\phi$ can now be described in terms of amplitude, $\phi_{\alpha}$,
and phase, $\theta_{\alpha}$, bosonic modes as
$
\phi_{R,L,\alpha}(y) = \phi_{\alpha}(y) \mp \theta_{\alpha}(y) .
$
In turn these bosonic fields can be written in terms
of charge and spin bosonic modes,
$
\phi_{\rho,s} = \frac{1}{\sqrt{2}} \left(\phi_{\uparrow} \pm
\phi_{\downarrow}\right)$
and
$
\theta_{\rho,s} = \frac{1}{\sqrt{2}} \left(\theta_{\uparrow} \pm
\theta_{\downarrow}\right)
$,
and it is easy to show that the Euclidean Lagrangean density
of the system can be
written as,
\begin{eqnarray}
{\cal L}_S = \sum_{i=\rho,s} \left\{\frac{g_i}{2 v_i}
\left[\left(\partial_{\tau} \phi_{i}\right)^2 + v_{i}^2
\left(\partial_y \phi_{i}\right)^2\right]\right\}
\label{ac}
\end{eqnarray}
$g_s$ and $g_{\rho} $are the Luttinger parameters
for spin and charge respectively and $v_s$ and $v_{\rho}$
their velocities of propagation.
Observe that this model exhibits
the phenomenon of charge and spin separation, as we can clearly see from
(\ref{ac}).

First let us assume that we are inside the antiferromagnetic
region and there is long range antiferromagnetic order ($\xi \to \infty$).
Let us consider the effect of the
stripe on the antiferromagnet. In this case the interaction
between the stripe and the antiferromagnet can be simplified
to a Ising-like form because the spin symmetry is broken (say
in the $z$ direction). That is, let us write
\begin{eqnarray}
{\cal H}_I = J' \sum_{m} S_{e}^z \left(S^z(+1,m) + S^z(-1,m)\right).
\label{kondo}
\end{eqnarray}
In the bosonized language $S_{e}^z = \partial_y \phi_s/\sqrt{\pi}$
which shows that only $\phi_s$ is coupled to the antiferromagnet.
It is easy now to trace the Luttinger liquid out of the problem
and calculate the effect of the Luttinger liquid on the spins
around the stripe. The final action reads,
\begin{eqnarray}
S = -\frac{(J')^2 v_s}{2 \pi g_s}\sum_n \int \frac{d k}{2 \pi}
\frac{k^2}{\omega_n^2 + v_{s}^2 k^2} |\Delta S^z(k,\omega_n)|^2
\nonumber
\end{eqnarray}
where
$
\Delta S^z(0,y) = S^z(0^+,y)+S^z(0^-,y)
$
is the total local spin around the stripe. Observe
that this action is retarded and non-local and the
limits of zero frequency and zero wavevector do not
commute in the integrand. If we impose the condition that the momentum 
is strictly conserved
and allow energy fluctuation (that is,
we take $k \to 0$ first), the interaction vanishes.
In the opposite limit ($\omega \to 0$ first),
if we require energy to be conserved while the momentum can
fluctuate, we obtain the physical interaction between the spins.
In the static case one obtains,
\begin{eqnarray}
S &=& - \frac{(J')^2}{2 \pi g_s v_s} \int dx \int d\tau
\left[ \left(S^z(0^+,y)\right)^2 + \left(S^z(0^-,y)\right)^2
\right.
\nonumber
\\
&+& \left. 2 S^z(0^+,y) S^z(0^-,y) \right] \, \, .
\label{ferro}
\end{eqnarray}
Therefore we have generated two kinds of terms: the first is a
single ion anisotropy at the border of the stripe and the second is
a {\it ferromagnetic} coupling between the opposite sides
of the stripe. This type of ferromagnetic coupling has been
already proposed in the literature in the context of magnetism
in the cuprates \cite{aharony}. The exchange energy is given by
\begin{eqnarray}
J_{eff} = \frac{(J')^2 a}{\pi g_s v_s}.
\nonumber
\end{eqnarray}
In order to estimate its value let us assume that the parameters
in the stripe are exactly the same as in the magnetic regions.
In this case $J' \approx J = 2 t^2/U$, $\tilde{t} \approx t$ and
$\tilde{U} \approx U << \pi v_F$ and one finds,
\begin{eqnarray}
J_{eff} \approx \frac{J}{\pi \sin\left(\frac{\pi}{2} n\right) \frac{U}{t} }
\nonumber
\end{eqnarray}
where $n = 2 (k_F a)/\pi$ is the electronic density on
the stripe. We see that $J_{eff} < J$ if
$\frac{U}{t} > \left(\pi \sin\left(\frac{\pi}{2} n\right)\right)^{-1}$
which is easy to attain in the antiferromagnetic
phase. This result implies that the coupling across the
stripes is weakened by the Luttinger liquid.

In a recent paper we have proposed an explanation for
the destruction of antiferromagnetism with hole doping 
in terms of the weakening
of the antiferromagnetic correlations around the stripe \cite{prl}.
This weakening introduces a spacial anisotropy in the propagation
of the spin waves parallel and perpendicular to the stripes.
The physical picture is that,
in the presence of stripes, the quantum fluctuations in the
system grow and at some critical value of the anisotropy
parameter a quantum phase transition occurs.

\begin{figure}
\epsfysize5cm
\hspace{1cm}
\epsfbox{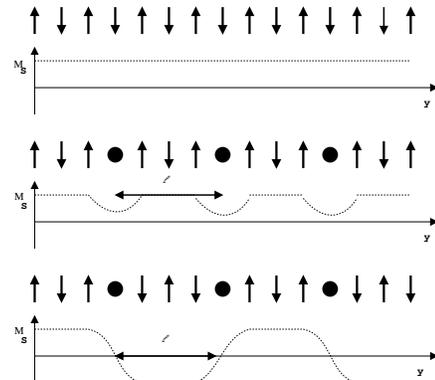}
\caption{Stagered magnetization as a function of position
for the possible configurations. From the top to the bottom:
homogeneous Neel state; Phase domain; Anti-phase domain.
Also shown is the interstripe distance $\ell$.}
\label{mag}
\end{figure}

Moreover, our results are consistent with the
neutron scattering data, as we explain now. It has been known for quite
some time that in La$_{2-x}$ Sr$_x$ Cu O$_4$ and related
compounds that instead of the magnetic peak at $(\pi/a,\pi/a)$
which is seen in the antiferromagnetic compound ($x \leq 0.02$)
one sees four incommensurate peaks at
$(\pi/a \pm \epsilon,\pi/a)$ and 
$(\pi/a,\pi/a \mp \epsilon)$ where $\epsilon$
depends on doping \cite{mason}. In some recent experiments by
Yamada {\it et al.} \cite{yamada}
this incommensurability effect was
measured with great precision. It was found that below
$x=0.05$ only one peak at $(\pi/a,\pi/a)$ is seen.
Above $x=0.05$ the four peaks are
observed and, in particular, for $0.06 \leq x \leq 0.12$
one finds $\epsilon$ to be proportional to the doping,
\begin{eqnarray}
\epsilon = \frac{2 \pi}{a} x
\label{extra}
\end{eqnarray}
where $a=3.8 A^o$ is the lattice spacing. Above $x=0.12$
$\epsilon$ seems to saturate. The four peaks in the
neutron scattering data can be easily explained in
the stripe picture by the formation of {\it anti-phase}
domains (see Fig.(\ref{stripe})). As is shown in Fig.(\ref{mag}),
the anti-phase domain leads to a change in the sign of
the staggered magnetization at the position of the stripes.
This change in sign implies a new scale in the problem which
is given by the interstripe distance $\ell$. It is easy to
conclude that,
\begin{eqnarray}
\epsilon = \frac{\pi}{\ell}.
\label{eps}
\end{eqnarray}

Thus equation (\ref{extra}) predicts that the interstripe
distance is decreasing with doping. This conclusion is
reasonable if one assumes that the doping does not
change the filling factor of the stripes. There are some
recent calculations using the exact solution of the Hubbard
model that seem to agree with the assumption \cite{chetan}.
It turns out, however, that the four peaks disappear below
$x=0.05$. That is, there is a commensurate-incommensurate
phase transition as a function of doping. Our results
indicate the nature of the phase transition. When
$\xi \geq \ell$ our picture is the one of a static stripe
with {\it ferromagnetic} coupling given by (\ref{ferro})
between the spins in
the neighborhood of the stripe. Therefore, in this regime
one expects a {\it phase domain} (see Fig.(\ref{stripe})).
It is clear that in this case the staggered magnetization
does not change sign at the stripe (see Fig.(\ref{mag})),
and one should find just one peak at $(\pi/a,\pi/a)$
as is seen in the experiment. Therefore we are led
to the conclusion that charge fluctuations must be
the source of the commensurate-incommensurate phase transition.
That is, the energy of the system is lowered by transverse
motion of the stripes. This result makes a lot of sense.
In the presence of a phase domain the transverse motion of
the stripe is frustrated (see Fig.(\ref{stripe})) because
of the increase in the antiferromagnetic energy due the hopping
of electrons with the same spin. This effect does not happen for the
anti-phase domain, because the hopping of a hole into the
antiferromagnet does not frustrate the spins (see Fig.(\ref{stripe})).
Therefore in an anti-phase domain the stripes can readily
fluctuate transversally.

\begin{figure}
\epsfysize6 cm
\hspace{0.0cm}
\epsfbox{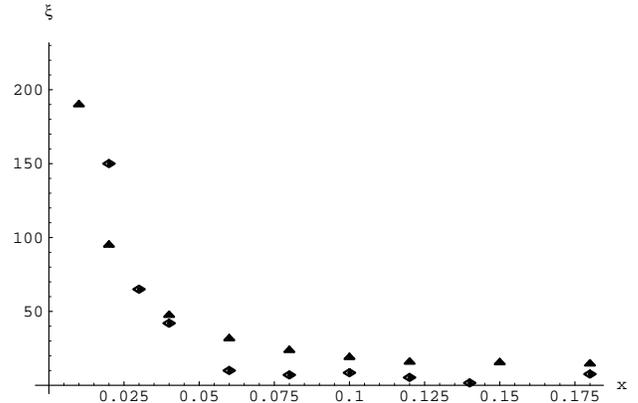}
\caption{Correlation length, $\xi$ (diamonds), from Birgeneau {\it et al.}
and Keimer {\it et al.} and interstripe distance, $\ell$ (triangles), from
Yamada {\it et al.} (all lengths are measured in $A^o$).
Notice the crossing around $3\%$ or $4\%$.}
\label{xi}
\end{figure}

Thus our conclusion is that when the stripes
start to fluctuate (which in our theory happens when
$\xi \approx \ell$) one should see this transition.
Indeed, in Fig.(\ref{xi}), one shows the experimental
data for $\xi$ and $\ell$ as a function of doping from two different
experiments. $\xi$ is obtained from Birgeneau {\it et al.} \cite{birgeneau}
and Keimer {\it et al.} \cite{keimer}
from the measurement of the width of of the magnetic peaks in the
doped samples. $\ell$ is obtained from Yamada {\it et al.} \cite{yamada}
from equation (\ref{eps}) for $x>0.06$ and
by extrapolating for $x<0.06$ using equation (\ref{extra}).
One clearly sees that the two sets of data cross around $x=0.03-0.04$.
We concluded therefore that the commensurate-incommensurate phase transition
can be associated with a phase to antiphase domain transition driven by
transverse fluctuations.

Let us look now at the effect of the antiferromagnet on the stripe.
We concluded that the coupling between the spins of the antiferromagnet
on opposite sides of the stripe is ferromagnetic. Therefore, in the ordered
phase, one must have 
$\langle S^z(+1,m) \rangle = \langle S^z(-1,m) \rangle = (-1)^m M_s$
where $M_s$ is the staggered magnetization. Substituting
this into (\ref{kondo}) one gets
\begin{eqnarray}
{\cal H}_I = 2 J' M_s \sum_{m} (-1)^m S_{e}^z(m)
\label{field}
\end{eqnarray}
which shows that the stripe sees a staggered magnetic field
in the $z$ direction proportional to $B(m) \approx - 2 J' M_s (-1)^m$.
One can easily show that the effect of this field is to
magnetize the Luttinger liquid (a paramagnetic effect).
The relevant part of the
Lagrangean can be rewritten in terms of the bosons as
\begin{eqnarray}
{\cal L} = \frac{g_s}{2 v_s}
\left[\left(\partial_{\tau} \phi_{s}\right)^2 + v_{s}^2
\left(\partial_y \phi_{s}\right)^2\right] + \kappa
\cos(\pi y/a) \partial_y \phi_{s}
\nonumber
\end{eqnarray}
where $\kappa = \frac{2 J' M_s}{\sqrt{\pi}}$.
Notice that this Lagrangean can be brought to a quadratic form
by a simple shift in the field $\phi_s$ as,
$
\Phi_s(y) = \phi_s(y) + \frac{a \kappa}{\pi g_s v_s} \sin(\pi y/a)
$
where the action for $\Phi_s$ is gaussian. It implies
that,
$\langle \phi_s(y) \rangle = - \frac{a \kappa}{\pi g_s v_s} \sin(\pi y/a)$.
It is easy to show that the mean value of magnetization on
the stripe is
\begin{eqnarray}
\langle S_z(y) \rangle = \frac{a}{\sqrt{\pi}} \langle \partial_y \phi_s \rangle =
- \frac{2 a J' M_s }{\pi g_s v_s} \cos(\pi y/a).
\nonumber
\end{eqnarray}
This gives a polarization of the stripe and now we
understand the origin of the ferromagnetic coupling
between the spins: an up spin in the magnetic region
polarizes a down spin cloud in the stripe which propagates
and couples with an up spin on the other side of the
stripe. This is just the zeroth order effect, because the transverse
degrees of freedom in the antiferromagnet are also coupled
to the Luttinger liquid, and from this argument it is not
clear what is their effect. For that we need a better study
of the transverse modes of the antiferromagnet. In the next
section we look at the problem coming from the disordered
phase and show that these modes actually play an important
role in the phase diagram of the stripe.

\section{$\xi > \ell$, magnetically disordered phase}

First let us turn our attention to the antiferromagnet.
The low energy physics of the
antiferromagnet can be described
in terms of the non-linear $\sigma$ model.
In order to treat this problem we go to a path integral formulation and
treat the Heisenberg spins in a spin coherent state representation.
The Euclidean
action associated
with the Heisenberg Hamiltonian, (\ref{Heisenberg}), is the usual
non-linear $\sigma$ model action that reads ($\hbar=1$) \cite{chn},
\begin{eqnarray}
S_H = \frac{1}{2 g} \sum_{i=R,L} \int d^2 r \int_0^{c \beta} d \tau
\left(\partial_{\mu} \hat{n}_i \right)^2
\label{sigma}
\end{eqnarray}
where $g=\sqrt{\chi_{\perp}/\rho_s}$ is the coupling constant,
$ \chi_{\perp} =4 J a^2$ the transverse spin susceptibility,
$\rho_s = J S^2$ the spin stiffness and $c = 2 S J a$ is the
spin wave velocity. $\hat{n}_R$ and $\hat{n}_L$ describe the
field $\hat{n}$ on the right and left sides of the stripe
respectively.
Observe that the imaginary time has been scaled by $c$.

We have concluded that the Luttinger liquid produces a ferromagnetic
coupling between the two sides of the antiferromagnet. This implies that
(\ref{kondo}) can be written as,
\begin{equation}
{\cal H}_I = J'S \sum_{m} (-1)^m \vec{S}_e(m) \cdot
\left(\hat{n}_R(+1,m)+\hat{n}_L(-1,m)\right).
\label{sigcoup}
\end{equation}
From (\ref{sigcoup}) one sees that the only field of
relevance for the electronic dynamics is the field
$\vec{u}(m,\tau) = \hat{n}_R(0^+,m,\tau)+\hat{n}_L(0^-,m,\tau)$.

In order to proceed one has to understand the dynamics of $\vec{u}$.
Here we use the well known
results for the non-linear $\sigma$ model. The partition function
of this model is given by,
\begin{equation}
Z = \int D\hat{n} e^{-S_H} \delta(\hat{n}^2-1)
\nonumber
\end{equation}
where $S_H$ is given in (\ref{sigma}) and the Dirac delta function
insures the norm of $\hat{n}$. This constraint can be introduced via
a Lagrange multiplier, $\lambda$ in the usual way,
\begin{eqnarray}
Z = \int D\hat{n} D\lambda e^{-S[\hat{n},\lambda]}
\nonumber
\end{eqnarray}
where
\begin{eqnarray}
S[\hat{n},\lambda] = S_H +i \int d^2 r \int_0^{c \beta} d \tau
\lambda \left(\hat{n}^2-1\right).
\nonumber
\end{eqnarray}
The constraint introduces non-linearities in this problem.
Here we treat this problem in a large $N$
expansion (where $N$ is the number of components of the vector $\hat{n}$)
\cite{largeN}. Since the symmetry is broken in the $z$ direction
we write $\hat{n} = (\sigma,\vec{\pi})$ where $\sigma$ is
the component of $\hat{n}$ in the $z$ direction and $\vec{\pi}$
are the $N-1$ transverse components. We can trace the
transverse modes out of the problem and the saddle point
equation when $N \to \infty$ can be obtained if we set
$\sigma=M_s$ and $\lambda= -i m^2/(2 g)$ in order to find
\begin{eqnarray}
M_s m &=& 0
\nonumber
\\
M_s^2 &=& 1- (N-1) g \sum_n \int \frac{d^2 k}{(2 \pi)^2}
\frac{1}{k^2+\omega_n^2+m^2} .
\label{meanfield}
\end{eqnarray}
The physical interpretation is straightforward: $M_s$ is the
order parameter (the staggered magnetization) and
$m$ is the inverse of the magnetic correlation length
$\xi$ ($m = 1/\xi$).
Therefore we have two phases: an ordered phase with
$M_s \neq 0$ and $m=0$ and a disordered phase
$M_s =0$ and $m \neq 0$ where $m$ and $M_s$ are given in
(\ref{meanfield}).
A major simplification happens in the disordered
phase. In this phase all the modes are massive ($m \neq 0$)
and therefore the physics of the problem is equivalent to a
{\it massive linear} $\sigma$ model \cite{sokol}.
That is, the theory is
essentially gaussian. In the disordered regime the action reads,
\begin{equation}
S[\vec{n}] \approx \frac{1}{2 g}
\int d^2 r \int_0^{c \beta} d \tau \left\{
\left(\partial_{\mu} \vec{n} \right)^2 + m^2 \vec{n}^2\right\}
\label{effac}
\end{equation}
where now the constraint is gone and $\vec{n}$ can vary arbitrarily.
This result leads to a major simplification.

The action (\ref{effac}) for the $\sigma$ model can be traced out
exactly because of its quadratic nature.
We proceed by introducing an identity on the partition
function as follows,
\begin{eqnarray}
1 &=& \int D\vec{u} \int D\vec{v} \exp\left\{-i \int dy \int_0^{c \beta} d \tau
\vec{v}(y,\tau) \cdot \left(\vec{u}(y,\tau)
\right. \right.
\nonumber
\\
&-& \left. \left.
\vec{n}_R(0,y,\tau)- \vec{n}_L(0,y,\tau)\right) \right\}
\label{trick}
\end{eqnarray}
where $\vec{v}$ is a Lagrange multiplier. Observe that
because the action in (\ref{effac}) is completely quadratic we
can integrate the fields $\vec{n}$ and $\vec{v}$ out and get an
action for the field $\vec{u}$ alone. We just quote here the
final result for the partition function for the fermions on the stripe
which has the form,
\begin{eqnarray}
Z = \int D a D a^* e^{- S_S[a,a^*] - S_I[a,a^*]}
\nonumber
\end{eqnarray}
where
\begin{eqnarray}
S_{I} = - \frac{3 g}{4} \left(\frac{J' S}{a c}\right)^2
\sum_n \int \frac{d k}{2 \pi}  \frac{|\vec{S}_e(k+\pi/a,\omega_n)|^2}{
\sqrt{k^2+\omega_n^2+m^2}}
\nonumber
\end{eqnarray}
and $S_S$ is the action associated with (\ref{ac}).
Observe that the interaction is a
retarded, temperature dependent and long range spin-spin interaction
which in real space becomes
\begin{eqnarray}
S_I = - \int dy d\tau dy' d\tau'
F(y-y',\tau-\tau') \vec{S}_e(y,\tau) \cdot \vec{S}_e(y',\tau')\
\nonumber
\end{eqnarray}
where,
\begin{eqnarray}
F(y,\tau) = \frac{3 g \nu \cos(\pi y/a)}{16 \pi a^4}
\sum_n e^{-i \omega_n\tau}
K_0(|y| \sqrt{\omega_n^2+m^2})
\nonumber
\end{eqnarray}
is the effective spin-spin interaction ($K_0$ is a modified
Bessel function). At zero temperature it becomes
\begin{eqnarray}
F(y,\tau) =  \frac{3 g \nu \cos(\pi y/a)}{24 \pi a^4}
\frac{e^{-m \sqrt{\tau^2+y^2}}}{\sqrt{\tau^2+y^2}}
\nonumber
\end{eqnarray}
where $\nu =J'/J$. Since we are interested in the propagator in real
time we have to Wick rotate from Euclidean space to Minkowski space
($ \tau \to -i c t$). The result is
\begin{eqnarray}
F(y,t) =  \frac{3 g \nu \cos(\pi y/a)}{24 \pi a^4}
\frac{e^{-m \sqrt{y^2-c^2 t^2}}}{\sqrt{y^2-c^2 t^2}}.
\nonumber
\end{eqnarray}
Thus for space-like separations ($y^2>c^2 t^2$) the propagador
decays exponentially at large space-like separations. But, for
time-like separations ($y^2<c^2 t^2$) we get an oscillatory
decaying function with two periods $\xi$ and $a$.

In the special case where the
correlation length goes to zero ($m \to \infty$, $\xi \to 0$) one gets,
\begin{equation}
F(y,\tau) = \frac{3 g \nu \xi}{16 a^4} \delta(\tau) \delta(y)
\label{local}
\end{equation}
and it becomes purely local. Taking (\ref{local})
and rewriting in Hamiltonian form one gets a simple renormalization
of the original Hubbard $\tilde{U}$.

At the other extreme case where the
correlation length diverges ($m=0$, $\xi \to \infty$)
the interaction reads,
\begin{eqnarray}
F(y,\tau) =  \frac{3 g \nu \cos(\pi y/a)}{24 \pi a^4} \frac{1}{
\sqrt{\tau^2+y^2}} \, \, .
\label{div}
\end{eqnarray}
The interaction is oscillatory and decays like $1/R$. It was
shown by Schulz that a $1/y$ interaction leads to Wigner crystalization
in a one-dimensional Fermi gas \cite{schulz}. This is exactly
the form of the instantaneous part of (\ref{div}) apart from
the oscillatory term.
Moreover, if we
take the zero frequency part of the interaction one finds that
$F(y,\tau) = \delta(\tau) U(y)$ where
\begin{eqnarray}
U(y)=\frac{3 g \nu \cos(\pi y/a)}{16 \pi a^4}  K_0(m |y|)
\nonumber
\end{eqnarray}
which for $\xi>>y>>a$ becomes,
\begin{eqnarray}
U(y)=\frac{3 g \nu}{16 \pi a^4} \cos(\pi y/a) \ln(\xi/|y|))
\nonumber
\end{eqnarray}
which is divergent in the antiferromagnetic phase $\xi \to \infty$.
Our conjecture is therefore that the system is crystalized in the
ordered phase. This result is
consistent with the experimental fact that the system is insulating
in this phase. We have reason to believe, therefore,
that the stripe undergoes
an insulator-metal transition with doping as the system goes
through the antiferromagnetic-paramagnetic phase transition. However, in
order to prove this conjecture we have to study how the
interaction renormalizes as we go to lower frequencies. This can
be done using renormalization group calculations with two frequencies:
the Fermi energy of the electrons on the stripe and $c/\xi$ which is
the characteristic energy of the spin waves.  Although calculations
with two frequencies have been done for electron-phonon interactions
\cite{twofre}, they have not been done for electron-magnon interactions.
One can show that this type of interaction leads to new diagrams
that are not considered in the electron-phonon problem which are related
to spin flip processes \cite{new}.

\section{Conclusions}

We have discussed here the limit of $\xi > \ell$ where
magnetism dominates over charge effects. In the opposite limit,
$\xi < \ell$, the physics is dominated by stripe
fluctuations and formation of anti-phase domains. In this
phase the stripes move transversally and clearly one expects
the transverse motion to be related to the existence of
superconductivity in cuprates \cite{zaanen,steve}. Indeed,
it is shown experimentally in La$_{2-x}$
Ba$_x$ Cu O$_4$ \cite{barium}
and in La$_{1.6-x}$ Nd$_{0.4}$ Sr$_x$ Cu O$_{4}$ \cite{tranquada}
that the pinning
of stripes reduces drastically the superconducting temperature
of the material. In the case of the model presented in this
paper it implies that the hybridization $V$ of the stripe
with the magnetic region becomes large and holes can hop
inside the magnetic regions. Moreover, one expects in this region
that interaction between the stripes will be important. In
particular, the stripes can exchange electrons (and Cooper
pairs) over the magnetic regions and the approximation of
an isolated stripe is not a sensible one. We believe however
that the main ingredients for the physics in this region are
already present in (\ref{mod}). It is worth noticing, however,
that most of the methods used in this paper can still be used
for $\xi < \ell$ if we make the appropriate changes \cite{new}.

In this paper, therefore, we focused on the limit where $\xi > \ell$.
We showed that in the ordered phase the effect of the Luttinger liquid on the
antiferromagnet is to generate a {\it ferromagnetic} coupling
between the spins around the stripe and lead therefore to the
presence of {\it phase domains} in contrast to the anti-phase
domains which are seen in the samples with doping $x>0.05$.
The physical reason for this ferromagnetic ordering is  
a spin polarization of the stripes.
We conjecture that the commensurate-incommensurate transition
seen in neutron scattering is associated with the phase to anti-phase
domain transition in the coupling between the spins. This transition
is driven by the gain in the kinetic energy of the transverse motion
of the stripes over the antiferromagnetic energy that tends to
suppress the transverse motion. Moreover, these results
are consistent with earlier work on the destruction of antiferromagnetism
in the striped antiferromagnet due to the growth of quantum fluctuations
\cite{prl}. Furthermore, in the disordered phase we have shown,
using a $\sigma$
model calculation, that close to the antiferromagnetic transition
the interaction between the electronic spins on the stripe becomes
singular, which leads to conjecture that the Luttinger liquid
condenses into a Wigner crystal. In this case the system is completely
insulating, in agreement with the experiments.

In summary, in this paper we propose a model for the physics of
stripes in
antiferromagnets in which the stripes are described by a Luttinger
liquid hybridized with a magnetic host. Using bosonization and
$O(N)$-$\sigma$ models we discuss the physics of
this problem in the limit where the physics is dominated by
the magnetic interactions, that is, $\xi > \ell$. We show that
it is possible with very general arguments to explain a series
of experiments in cuprates and to predict a transition in
the spin orientation with doping in these materials.

I am deeply indebted to Eduardo Miranda for many illuminating
phone conversations on stripes and other beasts. I also thank Daniel
Hone for a critical reading of the manuscript. Finally I
would like to thank Alexander Balatsky, Ward Beyermann,
Guillermo Castilla, Daniel Hone, Bernhard Keimer,
Steven Kivelson, Douglas MacLaughlin and Chetan Nayak
for helpful discussions.

\end{document}